# Fogging Jyaguchi Services in Tensai Gothalo

Gautam Bishnu Prasad [*1#2], Batajoo Amit [*1], Wasaki Katsumi [#2]

[*1#2] Department of Integrated Media, Wakkanai Hokusei Gakuen University Wakkanai, Hokkaido, Japan
[#2] Interdisciplinary Graduate School of Science and Technology, Shinshu University 4-17-1 Wakasato Nagano-city, Nagano, Japan

*Abstract*—This paper describes the efficient method of fogging in Tensai Gothalo. Tensai Gothalo is a novel dynamic router device developed in Gautam-Asami Laboratory of Wakkanai Hokusei Gakuen University which has sensing, actuating, monitoring and movable capability. Similarly, fogging is a new concept of cloud computing at which the data plane is defined in user device. In this paper we would like to present the stepwise explanation about how to fog in Tensai Gothalo. Furthermore, we will elaborate a technique to decentralize data with improvement in QoS and reducing latency without affecting the legacy services of clouds that can still work together while needed.

Keywords — *Cloud Computing, Jyaguchi, Tensai Gothalo, IOT.*

## I. INTRODUCTION

Fog Computing is a new paradigm that enhances the Cloud computing paradigm from data center plane to the clusters of end-user-devices plane which we termed fog in this paper. Cloud computing has shifted computing resources more or less from user plane to data center plane thereby centralizing the computing infrastructure into huge data centers. In contrast, fog computing decentralized the resources from cloud centers to the end-users or to the edge of the network, thus enabling a new breed of applications and services with newer potential. More specifically, fog computing is a computing paradigm that brings data processing, service utilization, networking, storage and analytics closer to the devices and applications that are more closer to the users. In this paper we argue that fog computing is a new paradigm of computing that uses the platform of Internet of Things (IoT), Smart Grid, Smart Communities or newly developed computing devices.

## II. SIGNIFICANCE OF FOG INFRASTRUCTURE

Due to intense growth of internet since few years back, computing resources is now more ubiquitously available. This growth has brought significance achievement for example it enabled the realization of a new computing concept called Cloud Computing. Cloud computing infrastructure is categorized into three main categories such as SaaS, IaaS, PasS. Though they seemed to demarcated into three different categories, all of these are utilized to build the application. The primary big giants of Cloud Computing are Google, Microsoft, and Amazon.

The main goal of Cloud computing is to leveraging the Internet to consume software or other IT services on demand. Cloud users can utilize and share processing power, storage space, bandwidth, memory, applications and software in number of ways. In response to their usage, cloud providers charges the users as per their consumption. This sort of business concept has been derived from the concept of utility business and thus cloud computing sometimes refers to utility computing too. In this way, users are not required to set up or to buy hardware by themselves as it used to be traditionally. It has brought a huge paradigm shift in the market. However, it has not addressed all issues raised in the user front.

Regardless of its supremacy in terms of providing resources to the end users, it has number of issues.

1) Cloud computing has arisen with new data security challenges[1], [2]. Existing data protection mechanisms such as encryption have failed in preventing data theft attacks, especially those perpetrated by an insider to the cloud provider.

2) A notable research to address this issue by applying disinformation attack method that returns large amounts of decoy information to the attacker has been introduced. However, once the data are transferred from your local LAN to the internet, there always remains the vulnerability of data theft.

3) The more significant issue often noted in literature is about data residency. For example, an issue[1] of data location has been pointed out. Cloud data center can cross the country or continent so that there would be a great difference between the rule, policy and the laws between the consumer society and provider's society. Furthermore, the issue of potential access to data by foreign governments is part of a wider issue, which is that the use of services based in other countries may result in customers being affected by laws of those countries.

In order to solve these kinds of issue, a new kind of computing architecture is necessary. The above mentioned data theft issue can only be minimized by keeping the data inside your premise. And, this can be offered by utilizing fog computing infrastructure

### III. RELATED RESEARCH

Research related to fog computing has been emerged since this term has been coined by Prof. Salvatore J. Stolfo originally. However, in an industrial research front, Cisco System Inc [3] seems to be the frontier company to adapt this technology. Being relatively a new term, there are very few researches that play significant roles into the literature. There are some research in academic area for example [4] is using fog computing infrastructure for health care system. Similarly,[5] [3] in focusing mobile users, whereas [6] highlights some security issues in fog computing.

Our research does however not only explore the existing methods and challenges faced by cloud but also conduct practical research and explore new methods that can enhance fog computing research. Furthermore, we have integrated fog infrastructure with newly developed devices such as Tensai Gothalo, the authors previously developed

### IV. DOMAIN OF JYAGUCHI FOG

We have identified the following areas where Jyaguchi fog can contributes its feature. We are highlighting the major characteristics that Jyaguchi fog should maintain, the main characteristics of the Jyaguchi Fog are highlighted as follows:

1) Should manage wide-spread geographical distribution such as for sensor nodes;
2) Should have support to mobility access
3) Should have diversity of devices, nodes or users
4) Should maintain essential access to varieties of link connections such as wired and wireless;
5) Should have strong presence of streaming and real time applications,
6) Should have heterogeneity of devices and data sources.
7) Should support of big data analysis.
8) Should have low latency for quality of service and location awareness.

### V. TENSAI GOTHALO AS AN INFRASTRUCTURE NODE OF FOG

Tensai Gothalo[7], [8] is a robotic vehicle that has routing and monitoring capabilities. This device can be used either as router device or as end-user devices. The cluster of which can also create fog infrastructure. In a current paradigm of Internet of Everything (IoE), we can take Tensai Gothalo as a thing. A thing in IoE is any natural or man-made object that can be assigned an IP address and participate in the communication. Whereas, a thing of Jyaguchi fog can transmit data over a Tensai Gothalo network which resides closer to the end-users. Transmitting data through internet to a data center might consume great deal of bandwidth. However, data transferred by Jyaguchi fog can be controlled whether to cross routers or put into the fog devices of Tensai Gothalo. Data resided in fog devices do not always cross the access router and thus reduce the suffering of latency.

### VI. ARCHITECTURE AND RESOURCE ALLOCATION DECISION PROCESS IN JYAGUCHI FOG

Fig. 1 shows the overall architecture of Jyaguchi fog infrastructure. Jyaguchi have adopted with the new phenomena of controlling the residency of services. Jyaguchi had often categorized its types of services in terms of granularity. It has the following categories:

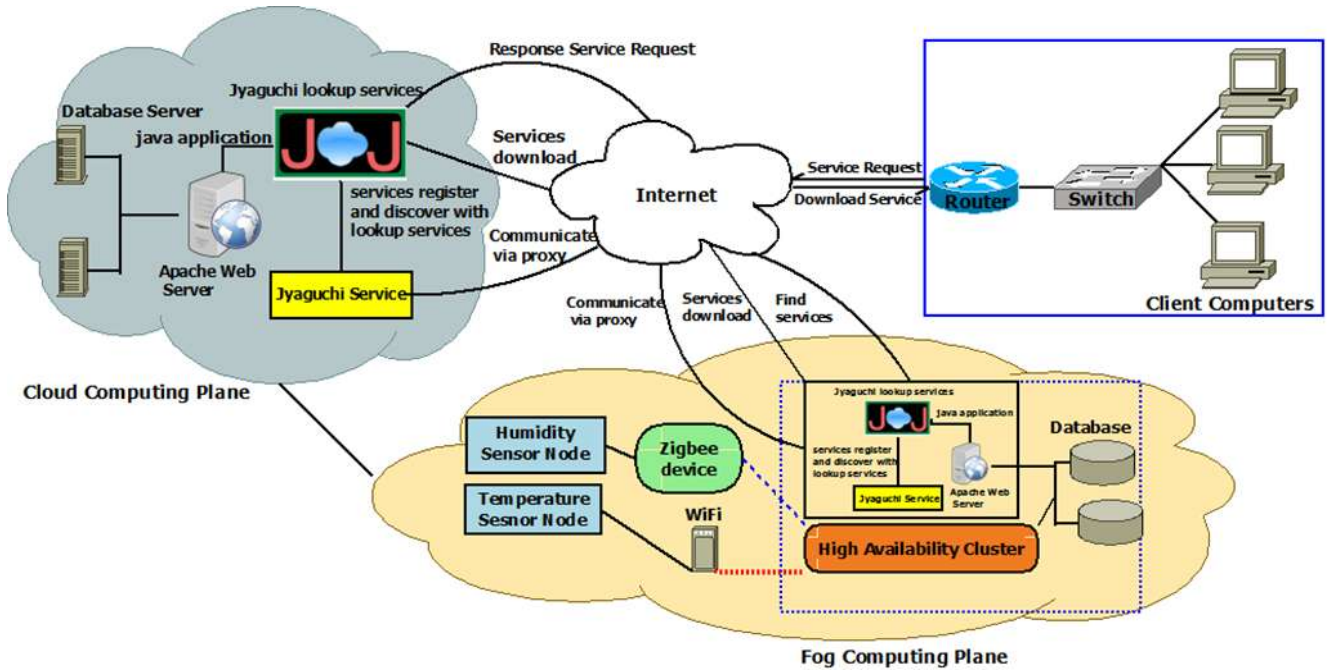

Fig. 1 Jyaguchi Fog Architecture

1) Mini Services: Services have very minimal granularity
2) Macro Services: Services relatively have larger granularity than mini services
3) Mega Services: Services with larger granularity than macro services and which are preferred to reside within end-user premises or relatively closer to end-users.

The details about the granularity is described in [9] detail. Among these services, we found out that few services are necessary to handle properly. For example, services which need higher security and the services which have higher granularities. Services that needs higher securities are vulnerable while they are exposed in the public cloud. Therefore, these sorts of services may require to keep inside the premises with higher securities. The other types of services are mega services which cost higher bandwidth are recommended to keep inside the premises. Furthermore, in case of keeping mega services outside the end-users premises, it is recommended to keep them as near as to the user's network. In such a case, multiple fog environments can be created thereby proving a cluster of Jyaguchi fog. A topology based redirection can be applied in order to redirects the traffic to the nearest users networks.

*A. Service Allocation Decision Process*

Resource sharing or allocation process is the process which tries to reduce the mean response time in order to access the resource in a network. A very simple line of thought to decide where to access depends upon the response time. The greater the time it requires to response, the lower the performance it offers. There needs a service allocation decision process that must account for the fact such that when designing an algorithm for resource sharing in the Jyaguchi fog, a quantifying method should be applied in order to measure the performance and the efficiency. First of all let's describe our consideration about quantitative evaluation approach for the decision process. In this evaluation process, we highly advocated the evaluation criteria by measuring the response time (RT) of the resource. RT would be core metrics to account its performance.

RT value could be the response time of entire cluster or a particular service. The equation of which is given as below:

Total Response Time = $\sum_{i=1}^{n}(each\ host(i))$ ----1

Equation 1 indicates the method of calculating total response time of the infrastructure. The higher the response time, the lower is performance. Response time can have influences of CPU, memory and the network bandwidth. The cumulative effect will be reflected to the value of response time.

Jyaguchi service provider can select the better RT value in which it can register its service. Whereas Jyaguchi client can access to the resource where the RT value is low. Considering this value, resource allocation policy can be defined more effectively.

The same formula will be enhanced in order to recommend the metric for which the client and service provider can take the reference. This reference values can be taken automatically through program or by manual.

Priority Index (PI) = (Total RT)/(Total Number of Measurement)

Where the value of PI is obtained by dividing the total response time by the total number of measurement. This is the mean value of the response time which will be taken into consideration while allocating the resource. Jyaguchi service provider or user can state this PI metric with the representation as shown in figure 2. The advantage by following this approach is that it does not require spending large amounts of resources or time in client side to thoroughly investigate what services are performing well and actively available in the fog. This approach can be utilized in order to exploit it fully in a reliable manner.

In figure 2, an example of visualization map representing the priority indicator is shown. Where the blue colored nodes indicates that the best recommended candidate for which both the services provider or the client can utilize either to access the resource or to publish the resource. Similarly, yellow colored needs higher caution to use whereas the red ones are not recommended at all.

## VII. IMPLEMENTATION OF FOG SERVICES

Conventionally Jyaguchi platform has been used in order to develop cloud computing services. Furthermore the platform of Jyaguchi has also been utilized in hybrid computing infrastructure network integrated with smart devices and Micro Engineering Tools [9] in cloud. Micro Engineering Tools are the highly dynamic and interactive services developed in the Jyaguchi Cloud platform at which Java based application can be built and exported to the client over a network[10]. The implementation process is quite similar with previously implemented process. Thus, there is no overhead in learning to adapt the new computing infrastructure. We have already discussed the significance of fog computing infrastructure in section 2. In this section we discuss the implementation process of fog services. For our description, we consider only few services which utilizes our premises in order to collect the real time data. For example, we have built upon weather report services by using our own hardware units. These units are equipped with sensor nodes which collect the data from the surrounding environments. For our experimental setup we have utilized temperature and humidity sensors. The figures of these sensors, circuit and the communication with Raspberry pi is also given in the figure 4. The basic reason of the development of Fog Services in Tensai Gothalo [7], [11] is to improve efficiency of data to be transported in fog network for data processing, analysis and storage. As previous studies and the researches contributed to emerge Tensai Gothalo as one of the reliable and robust network devices, we selected fogging infrastructure to build upon the services integrated with this device. Next, let's discuss the process of fogging Jyaguchi in Tensai Gothalo in stepwise manner.

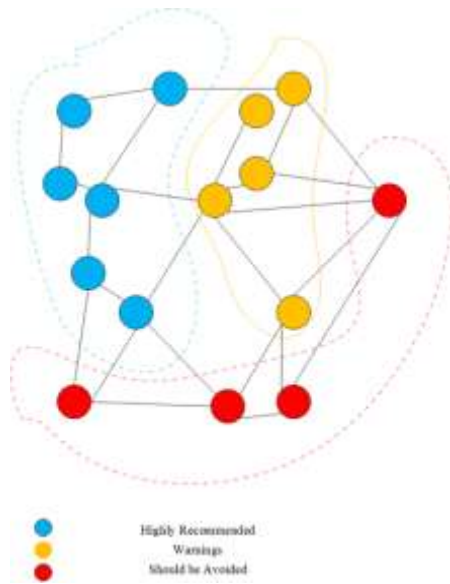

Fig. 2 Visualization Map of Priority Indicator

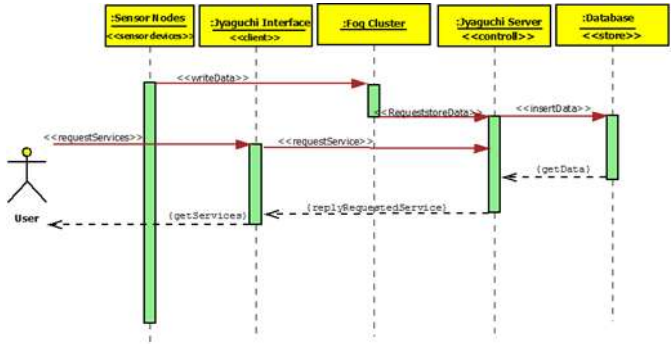

Fig. 3 Standard Sequence Diagram of Fog Services

Phase 1: First of all we need to setup fog infrastructure. This infrastructure needs different components. The details of our experimental setup developed in our lab is provided in table 2.

Table 1: Specification of Fog Computing Experimental Setup

| Name of Device | Details | Quantity |
|---|---|---|
| Tensai Gothalo | Master | 2 |
| Tensai Gothalo | Slave | 4 |
| Cluster | HA Proxy | 1 |
| Sensors | Humidity/temperature and other sensors (**e.g DHT22/DH11**) | As per requirement |
| Power Backup System | • Poggimo device<br>• Solar Panel and Battery | 2 |
| Other network devices | Router/Switch | As per requirement |

Phase 2: Next we need to install Jyaguchi infrastructure in TG. However TG are required to

Table 2: Details of Jyaguchi Computing Resources

| Resources | Details | Remarks |
|---|---|---|
| Lookup service | Apache River 2.2.2 | For unicast and Multicast lookup service |
| Service Provider | Jyaguchi Platform | Micro, Macro and Mega Services |
| Database server | MySQL Database server | Version no: 5.5.43 |
| Web server | Apache | Version No:2.2.22 |
| Language | Java | |
| Client Interface | Jyaguchi Universal Browser | |

equip with sufficient computing resources. This has been achieved by integrating Raspberry PI in TG. The details of Jyaguchi Infrastructure is given in table 2. As we are utilizing Jyaguchi infrastructure,

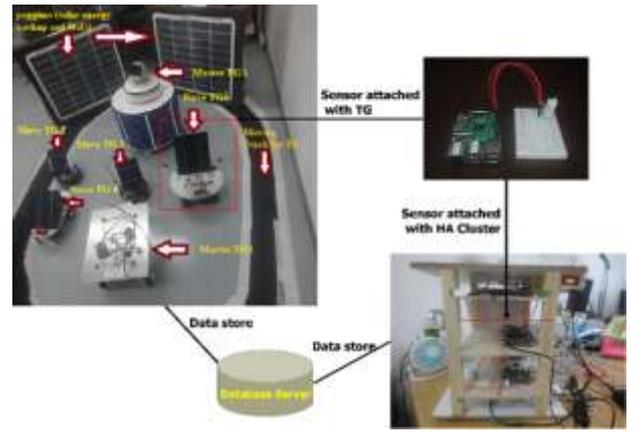

Fig. 4 Experimental setup of hardware

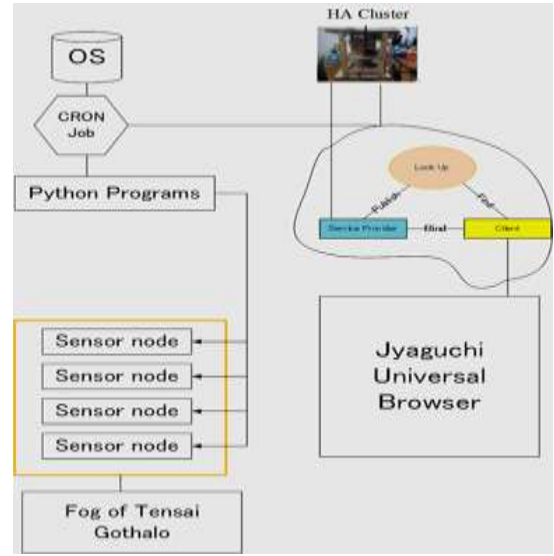

Fig. 5 Sensor Nodes and Jyaguchi System

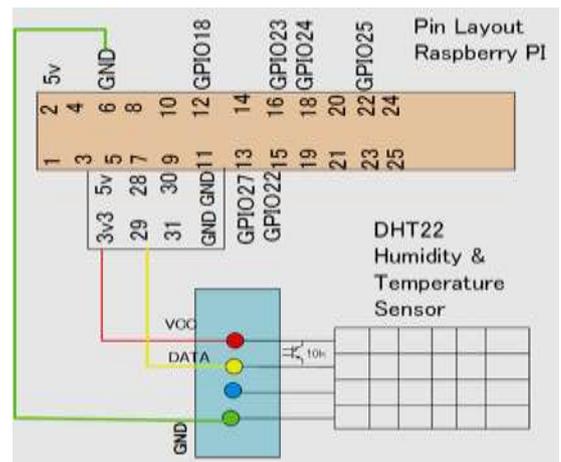

Figure 6: DHT22 Sensor Integration with Raspberry PI

the implementation of Jyaguchi services are required to follow its convention. Jyaguchi utilize

SOA while developing the services. It involves lookup service[9], [10], [12] that provides the platform for service registration and client lookup so that client can find the location information of the service.

Phase 3: In this pahse, the real development of the services is done. As we follow the same strategies Development of fog services in Jyaguchi platform.

### A. Temperature Service

This service is implemented in order to measure the temperature of the surrounding environment. We used AYWS DHT22/DH11 sensor which can be bought relatively in a low cost than other sensors, and can be activated by using 3-5 v in order to build this service. Easy to install however a careful handling is necessary if the surrounding environment is full of chemical vapor as it can hampered the sensor and its calibration to produce accurate data. This sensor is connected with Raspberry PI. Sensor produces digital data which are pruned with python script and the intended data of temperature has been passed to the database server. This data is retrieved and the data is plotted by temperature service. A data are represented by using our temperature service as shown in figure this service can be downloaded by users.

### B. Humidity Service

Similarly, the data of humidity in the air has been sensed by this sensor and this data is also plotted by Jyaguchi service. The figure of which is shown in figure 6. As Jyaguchi services are built upon SOA based architecture [10] users within the fog or cloud can utilize it when required.

### C. Migration of Legacy Services

Most of the time, it becomes challenging task to cut over legacy services to targeted services that can adapt new environment. A through study is required to complete migration task. We need to consider the capacity of CPU, memory power, software platform and other dependencies. This task includes varieties of techniques such as reverse engineering, business logic reengineering, schema mapping and translation, data transformation, application maintenance, human computer-interaction and testing. There are few techniques introduced in the literature[13], [14]. However, the convenient method of migration is to wrap the legacy services with new interface that can operate as a proxy module which can fit new environment. Also following the step wise converting approach from legacy to new service is recommended. The limited computing resources as experimented in our test case, may generate certain problems while migrating the legacy services. We have successfully migrates our legacy services such as calculating service, editor services and few other services which were previously developed for the cloud computing environment.

### VIII. EVALUATION AND RESULTS

We have developed micro services which were entirely depended upon low level hardware. We also needed to activate sensor and interpreted its sensing data which requires low level language. In order to communicate with the device output, we utilized python script the result of which was inserted in database. These data are re-queried by Jyaguchi services and displayed accordingly. Data are polled in every minute basis as every second will generate un-necessary load for the device. The precisjons of humidity and temperature entirely depend upon the quality of sensor hardware.

Our results [figure 7] shows that cut over of legacy systems and the development of new services in Tensai Gothalo is possible however certain QoS may suffer if the computing capacity of Tensai Gothalo is not expanded for example by clustering the hardware resources or by using virtualization technology. The research of which is need to be explored further.

### IX. FUTURE WORKS

We have already built up Jyaguchi fog infrastructure which can communicate with our sensor nodes some which are carried by our Tensai Gothalo. Future works will emphasize to build upon service dispatcher that can publish the service as per the security requirements of the services. We would also like to implement following services in the future

1) Other geographical environmental data such as barometric pressure, wind speed, rain fall, sun light exposure, GPS value and many others which are important for particular geographical areas.

2) A meaningful data analysis services so that each sensor data can be utilized by the users with meaningful interpretations.

3) A service dispatcher that can dynamically allocate the service either in the fog or in the cloud without requiring the interruptions of the service provider.

## X. CONCLUSION

In this paper, we pointed out some challenges of cloud computing sector which can be addressed by fog computing infrastructure. There are millions of users who are worried about data theft that can be happened in cloud infrastructure due to the issue of data residency. However, in fog computing infrastructure, once the computing premises is domiciled in the end user plane, it is relatively safer. Furthermore, the problem of insider data theft attacks which can be happened more in case of cloud can be lowered in fog computing infrastructure by dynamically generated decoy files. The reason of worries in cloud is that threat of malicious attacks are happening due to lack of transparency in cloud providers. In order to overcome these kinds of fears pervaded among users, we recommend to utilize fog infrastructure for the services which can be deployed in such infrastructure. To this effect, we have successfully demonstrated the experimental setup of fog computing infrastructure supported by Tensai Gothalo in which fog services can be built upon. One of the big challenge adopting fog computing infrastructure is that while you cut over legacy system into a new system a targeted new system must be able to retain the functionality and the quality, and the important data of the original legacy system which were successfully obtained while fogging Jyaguchi in Tensai Gothalo.

### ACKNOWLEDGMENT

We would like to thank to the entire member of Gautam-Asami Seminar who directly or indirectly help for our research. Also thanks reserve by the staff of Wakkanai Hokusei Gakuen University and Shinshu Unviersity for their continuous support.